\documentclass[12pt]{iopart}

\expandafter\let\csname equation*\endcsname\relax
\expandafter\let\csname endequation*\endcsname\relax

\usepackage{etoolbox}
\usepackage{ulem}
\makeatletter
\def\@mkboth#1#2{}
\newlength\appendixwidth
\preto\appendix{\addtocontents{toc}{\protect\patchl@section}}
\newcommand{\patchl@section}{%
  \settowidth{\appendixwidth}{\textbf{Appendix }}%
  \addtolength{\appendixwidth}{1.5em}%
  \patchcmd{\l@section}{1.5em}{\appendixwidth}{}{\ddt}%
}
\makeatother

\usepackage{cite}
\usepackage[colorlinks, citecolor = cyan, urlcolor = blue]{hyperref}

\usepackage{epsfig}
\usepackage{latexsym}
\usepackage{bm}

\usepackage{graphicx}

\usepackage{tikz}
\usetikzlibrary{positioning, patterns}

\usepackage{amsmath}
\usepackage{amsfonts}
\usepackage{amssymb}
\usepackage[mathscr]{eucal}

\graphicspath{{figures/}}

\usepackage{xcolor}

\usepackage{color}
\definecolor{dgreen}{rgb}{0,0.7,0}

\def\beq{\begin{equation}}
\def\eeq{\end{equation}}

\def\nn{\nonumber}

\begin{document}

\title[Crystal to liquid cross-over for trapped interacting active particles]{Crystal to liquid cross-over for active particles with inverse-square power-law interaction}
\author{Saikat Santra$^1$, L\'eo Touzo$^2$, Chandan Dasgupta$^{1,3}$, Abhishek Dhar$^1$, Suman Dutta$^1$, Anupam Kundu$^1$, Pierre Le Doussal$^2$, Gr\'egory Schehr$^4$, Prashant Singh$^5$\\}

\address{International Centre for Theoretical Sciences, Tata Institute of Fundamental ~Research, Bengaluru 560089, India$^1$}
\address{Laboratoire de Physique de l’Ecole Normale Sup\'erieure, CNRS, ENS and PSL Universit\'e, Sorbonne Universit\'e,
Universit\'e Paris Cit\'e, 24 rue Lhomond, 75005 Paris, France$^2$}
\address{Department of Physics, Indian Institute of Science, Bangalore 560012, India$^3$}
\address{Sorbonne Universit\'e, Laboratoire de Physique Th\'eorique et Hautes Energies,
CNRS UMR 7589, 4 Place Jussieu, 75252 Paris Cedex 05, France$^4$}
\address{Niels Bohr International Academy, Niels Bohr Institute,
University of Copenhagen, Blegdamsvej 17, 2100 Copenhagen, Denmark$^5$}
\ead{saikat.santra@icts.res.in, leo.touzo@phys.ens.fr}
\date{\today}

\begin{abstract}
\noindent
{We consider a one-dimensional system comprising of $N$ run-and-tumble particles confined in a harmonic trap interacting via a repulsive inverse-square power-law interaction. We numerically compute the global density profile in the steady state which shows interesting crossovers between three different regimes: as the activity increases, we observe a change from a density with sharp peaks characteristic of a crystal region to a smooth bell-shaped density profile, passing through the intermediate stage of a smooth Wigner semi-circle characteristic of a liquid phase.   
We also investigate analytically the crossover between the crystal and the liquid regions by computing the covariance of the positions of these particles in the steady state in the weak noise limit. It is achieved by using the method introduced in Touzo {\it et al.} [Phys. Rev. E {\bf 109}, 014136 (2024)] to study the active Dyson Brownian
motion. Our analytical results are corroborated by thorough numerical simulations.}
\end{abstract}

\makeatletter

\tableofcontents

\newpage
\section{Introduction} 
Active particles refer to a class of non-equilibrium systems where every individual unit breaks detailed balance condition~\cite{marchetti2013hydrodynamics,bechinger2016active,Ramaswamy_2017}
. Powered by an internal energy source, active particles exhibit stochastic but directed motion. Even understanding steady state properties becomes nontrivial and one has to go beyond equilibrium statistical physics.  A notable example is the run-and-tumble particle (RTP) which is inspired from the motion of certain bacteria like E. Coli~\cite{berg2004coli,Tailleur_PRL_2008}. At the single particle level, striking properties have been uncovered in the form of steady states \cite{Dhar_PRE_2019,sevilla2019stationary}, first passage properties~\cite{angelani2015run,gueneau2023optimal}, extreme value statistics and the convex hull problem~\cite{hartmann2020convex,singh2022mean}, which show interesting and notable distinctions from passive Brownian motion~\cite{Malakar_JSM_2018,Dhar_PRE_2019,Basu_JPA_2020,Mori_PRL_2020,Singh_JPA_2022}. 

The exploration of interacting active particles has unveiled fascinating behaviors like flocking~\cite{TONER_AOP_2005}, jamming~\cite{Slowman_PRL_2016}, absence of the equation of state for pressure~\cite{Solon_nature_2015}, motility-induced phase separation~\cite{Fily_PRL_2012, Cates_ARCMP_2015} and universal properties in single file motion \cite{Dolai2020}. These features are otherwise absent in their passive counterparts. Recent works have initiated analytical studies into the dynamics of interacting active particles. However, there exist only few exact results even in one spatial dimension~\cite{Thompson_JSM_2011, Put_JSM_2019,Das_JPA_2020,Dandekar_PRE_2020,Singh_JPA_2021,Mukherjee_scipost_2023,Metson_EPL_2023,Metson_PRE_2023,touzo_arxiv_2023,Santra_JSM_2024}.

Very recently, a number of works have obtained analytical results for systems of RTPs with long-range interactions~\cite{Tanmoy_PRE_2024,Tanmoy_PRE_2024_2,Touzo_EPL_2023,touzo_arxiv_2023}. In particular, the active version of the Dyson Brownian motion (DBM) was studied extensively~\cite{Touzo_EPL_2023,touzo_arxiv_2023}. The DBM~\cite{Dyson_JKP_1962,Dyson_JMP_1962} is one of the well-studied models which shares a direct connection to the random matrix theory~\cite{forrester_PUP_2010,mehta2004random}. This model has found applications in several other areas such as quantum chaos~\cite{mehta2004random,Gerbino_arxiv_2024} and stochastic growth~\cite{prahofer2002scale}. In the DBM, the particles experience repulsive logarithmic interactions (i.e., a force $\sim 1/x$, with $x$ being the inter-particle distance) and are subjected to a global harmonic trap. The average density profile has the famous Wigner semi-circle (WSc) form 
~\cite{Dyson_JMP_1962}. Surprisingly, in the active DBM, across a broad range of the activity parameter $v_0$, it was found that the density profile in the steady state maintains the shape of the Wigner semi-circle, as in the passive DBM~\cite{Touzo_EPL_2023}.  Some analytical arguments were provided based on the computation of the covariance of the particle positions~\cite{touzo_arxiv_2023}.

Another interesting example of an interacting one-dimensional system is the case where the interparticle potential is of the inverse-square power-law form, which we  will refer to   as the inverse-square-model (ISM). In the context of Hamiltonian dynamics, such interaction has been well studied and is known in the literature as the Calogero-Moser model~\cite{Calogero_JMP_1969,Calogero_JMP_1971,Calogero_LNC_1975,MOSER_AIM_1975}. For the Hamiltonian case, this is an integrable system for which one can explicitly construct the integrals of motion using the Lax formalism~\cite{bogomolny_PRL_2009}. Systems with such inverse square interaction have also  appeared in various other  
physical contexts such as random matrix theory and soliton physics~\cite{Calogero_JPS_Polychronakos,Kulkarni_JPA_2017,bogomolny_PRL_2009}. {A surprising connection to the DBM is that even though the interaction potential of the inverse square law in ISM  and logarithmic in DBM are very different, it turns out that they share an identical minimum energy configuration which are given by the zeros of the Hermite polynomial $H_N(x)$ (see Refs.}~\cite{ForresterRogers1986,Agarwal_JSP_2019}). Hence they share the same WSc density profile in the large $N$ limit~\cite{Agarwal_PRL_2019,Agarwal_JSP_2019}. In addition, there are remarkable relations between their Hessian matrices, which characterize small oscillations around the minimum of the potential~\cite{Agarwal_JSP_2019}. In the present work we explore the connections between the two models once we introduce activity, in particular when the Brownian motion of individual particles is replaced by RTP dynamics.  
The active DBM was studied in Ref.\cite{Touzo_EPL_2023} and here we report and compare corresponding results for the active ISM. The main properties that we investigate are: (i) the mean density profile in different regimes of activity; (ii) the fluctuations of the particle positions in the limit of weak noise and their connection with the properties of the Hessian matrix. These fluctuations are used to quantify the observations of the `crystal' to `liquid' crossover in this system.

The remainder of the paper is structured as follows: Sec.~\ref{section_model} is devoted to the description of the model and the observables we are interested in. We present numerical results on the steady-state density profile and on its scaling with the system size in Sec.~\ref{section_density}. We find an interesting crossover: as the activity increases, the density profile undergoes a notable change from a multi-peaked crystalline structure at low activity, to a liquid-like state exhibiting a smooth Wigner semi-circular density profile at intermediate activity. This crossover is quantified through the computation of the Lindemann's ratio. As the activity increases further, the system undergoes another interesting crossover where the steady-state density profile changes from the Wigner semi-circle to a bell-shaped form. Following the approach developed in Ref.\cite{touzo_arxiv_2023}, we compute analytically in Sec.~\ref{section_hessian} the covariance of the particle positions and of the gaps in the limit of small activity and large persistence time, providing support for the observed transition in the density profile. Finally, we conclude with intriguing perspectives for future research in Sec.~\ref{section_conclusion}. Some additional details are provided in the Appendices.

\section{Model, observables and summary of main results} 
\label{section_model}
We consider the motion of $N$ run-and-tumble particles (RTPs) in one dimension, mutually interacting via a long-range repulsive inverse-square power-law potential of strength $g~(>0)$. In addition, the particles are subjected to a global confining harmonic potential with stiffness parameter $\lambda~(>0)$. Denoting the position of the $i$-th particle at time $\tau$ by $y_i(\tau)$, the time evolution equation is given by
\beq
\frac{dy_i}{d\tau}=-\lambda y_i+\sum_{\substack{j=1\\j \neq i}}^{N} \frac{2 g}{(y_i-y_j)^3}+ \bar{v}_0 \sigma_i(\tau),~~\text{for }i=1,2, \cdots, N,
\label{eqn_of_motion}
\eeq
where $\bar{v}_0$ is the speed of the particle and $\sigma _i(\tau)$ is the dichotomous noise that switches between $\pm 1$ at a constant rate $\bar{\gamma}$. Because of the infinite repulsion on contact,  the particles remain ordered, $y_1 <y_2< \ldots <y_N$ \cite{touzo_arxiv_2023}. For any pair of particles, the noises $\sigma _i(\tau)$ on different particles are statistically independent and their temporal correlation takes the form $\langle  \sigma _i(\tau) \sigma _j(\tau') \rangle = \delta _{i,j}  \exp \left( -2 \bar{\gamma} |\tau-\tau'|\right)$ with $\delta _{i,j}$ being the Kronecker delta. Note that for $\bar{\gamma} \to \infty,~\bar{v}_0 \to \infty$ with the ratio $\bar{v}_0^2/\bar{\gamma}$ fixed, the model reduces to the passive limit of Brownian particles with diffusion coefficient $D=\bar{v}_0^2/2 \bar{\gamma}$. However, for any other values of these parameters, we anticipate to see departures from the thermal case. It is instructive to write the equations in dimensionless form. Defining the dimensionless time and position variables $t=\lambda \tau$ and $x_i=y_i/(g/\lambda)^{1/4}$, the above equations take the form
\beq
\frac{dx_i}{dt}=- x_i+\sum_{\substack{j=1\\j \neq i}}^{N} \frac{2}{(x_i-x_j)^3}+ v_0 \sigma_i(t),~~\text{for }i=1,2,..N,
\label{eqn_of_motion}
\eeq
where now the only remaining dimensionless parameters are $v_0=\frac{\bar{v}_0}{\lambda^{3/4} g^{1/4}}$ and the tumbling rate $\gamma=\bar{\gamma}/\lambda$. {Note that, the activity in the system can be increased either by enhancing the speed $v_0$ or reducing the tumbling rate $\gamma$, although the two parameters have quantitatively different effects.  

In this work, we will be interested in steady-state properties. In particular, we consider the mean density of particles defined as follows
\beq
{\rho}(x)=\frac{1}{N} \sum_{i=1}^{N} \langle \delta (x_i-x) \rangle,
\label{density_definition}
\eeq
where $\langle ...\rangle$ denotes an average over the steady-state distribution. 

In our simulations we observe that increasing the activity in the system causes the density profile to cross over from a multi-peaked crystalline structure to a smooth Wigner semi-circle, and eventually to a bell-shaped profile. To characterize the first crossover from the crystal-like structure to the liquid-like structure, we examine the fluctuations in the positions of particles in the steady state. Therefore, we measure the mean positions, the variances, and the mean interparticle separations as follows
\begin{align}
\bar{x}_i&= \langle x_i \rangle,\\
s_i^2&=\langle ( x_i-\bar{x}_i) ^2\rangle \\
 \Delta_{i,n}&=   \bar{x}_{i+n}-\bar{x}_i.
\end{align}
Additionally, we compute the covariance of the particle positions which eventually leads to the computation of the variance of inter-particle gap 
\beq
g^2_{i,n}=\langle \left (x_{i+n}-x_{i} \right )^2 \rangle - \Delta_{i,n}^2.
\eeq Specifically, to study the crossover more quantitatively, we compare the standard deviation of particle positions to the average interparticle separation, known as the Lindemann's ratio,
\beq
\eta_i=\frac{s_i}{\Delta_{i,1}}.
\label{eq:lindemanns_ratio}
\eeq
To address the second crossover, from a smooth Wigner semi-circular density profile to a bell-shaped profile, we quantify the difference between the density profile in the steady state and the Wigner semi-circular profile $\rho_{\rm sc}(x)$ by computing the following distance measure
\beq
\chi=\int_{-\infty}^{\infty} |\rho(x)-\rho_{\rm sc}(x)| dx,
\label{eq:def_chi}
\eeq
where 
\beq
\rho_{\rm sc}(x)=\begin{cases}
 \frac{1}{\pi N} \sqrt{2N-x^2} \qquad \text{for} \qquad |x| \leq \sqrt{2N} \;, \\
 ~~~~~~~0~~~~~~~~~~~~~\;\;\text{otherwise}.
 \end{cases}
\eeq

In the following sections, we will numerically study the form of ${\rho}(x)$, the position fluctuations $\{s_i\}$, Lindemann's ratio $\{\eta_i \}$, the distance measure $\chi$ and the variance of inter-particle gap $g^2_{i,n}$ upon varying  (i) the activity parameters $v_0$ and $\gamma$ and (ii) the number of particles $N$.

It is useful to summarize our main results: 
\begin{enumerate}
\item As observed in the active DBM, the steady-state density profile in the active ISM also exhibits three distinct structures depending on the strength of the activity. At very small activity, the system resembles a crystal having a multi-peaked density profile. The Wigner semi-circle gives the envelope of the density profile (see Fig.~\ref{figure_density_large}(a)). As activity increases, the crystalline structure in the density profile vanishes, resulting in a smoother profile. This profile matches well with the Wigner semi-circle across nearly the entire support region, except at the edges (see Fig.~\ref{figure_density_large}(b)). The support of the density profile can be approximated as $[-\sqrt{2N}, \sqrt{2N}]$. By increasing the activity, the steady-state density profile deviates significantly from the Wigner semi-circle, transforming into a bell-shaped profile with support approximately on $[-v_0, v_0]$ where $v_0 \sim \mathcal{O}(N)$ (see Figs.~\ref{figure_density_large}(c) and ~\ref{figure_edge_scaling}). However, the scaling form described in Eq.~\eqref{eq:scaled_density} remains valid for the vast majority of particles. Interestingly, we observe that near the edges, the density profile decays following a power-law of the form \(\sim 1/x^3\) (see Fig.~\ref{figure_edge_scaling}). This power-law decay seems to be universal as long as the particles confined in a harmonic trap experience infinite repulsion upon contact. 

\item These three different structures in the density profile correspond to three different activity regimes, which are named as weakly active, intermediate and strongly active regime. These regimes are further quantified by the computation of the Lindemann's ratio $\eta_i$ (defined in Eq.~\eqref{eq:lindemanns_ratio}) and the distance measure $\chi$ (defined in Eq.~\eqref{eq:def_chi}). In the limit of small $v_0$, the particles exhibit only small fluctuations about their equilibrium configuration which actually leads to the density profile having multiple peaks. With increasing $v_0$, the fluctuations of the particles increase. Whenever the typical fluctuation of particle positions becomes comparable to the average separation between nearest neighbors, one expects to have a smooth density profile. The first crossover from a multi-peaked density profile to a smooth semi-circle density profile is captured by Lindemann's ratio, which measures the standard deviation of particle positions with respect to the average separation between successive particles. On the other hand, the second transition, from the Wigner semi-circle to bell-shaped profile is quantified by the computation of the quantity $\chi$ as defined in Eq.~\eqref{eq:def_chi} which measures the difference between the density profile $\rho(x)$ and Wigner semi-circle. Based on the numerical values of these two quantities $\eta_i$ and $\chi$, we draw a `phase' diagram in Fig.~\ref{fig:phase_diagram_together} showing three activity regimes for two different system sizes.

\item We use an analytic 
approach developed in Ref.\cite{touzo_arxiv_2023} based on small displacements around the ground state (i.e., in the limit $v_0 \to 0$) to describe the crossover from the `crystal' to the `liquid' region. Surprisingly, the predictions of the small displacement theory for the variance of the particle positions and gaps are quite accurate both in the `crystal' and `liquid' phases. In the large $N$ limit, we derive a closed expression for the variance of the particle positions in the small tumbling rate limit, as presented in Eq.~\eqref{eqn:mathcal_vb}. 
By comparing the typical fluctuations of the particle positions with the typical interparticle distance, we find a criterion for the first crossover, which is expected to occur when \(v_0 \sim \mathcal{O}(1)\), independent of the system size. On the other hand, by comparing the typical position fluctuations with the total support of the density profile in the passive limit, we find that the second transition is expected to occur at \(v_0 \sim \mathcal{O}(N)\). Additionally, from the covariance of the particle positions, we have computed the variance of the gap between any two particles. In the large $N$ limit, we derive an analytical formula for the variance of the interparticle gap \(g^2_{i,n}\) in Eq.~\eqref{eq:var_gap_final_form} near the center of the trap (\(i = N/2\)). All the analytical results are verified numerically. An extension to finite $\gamma$, based on a companion paper \cite{Riesz2024}, is also presented, and compared to numerical results.
\end{enumerate}

\section{Density profile}
\label{section_density}
In this section, we discuss the effect of activity on the average density profile in the steady state. For $v_0=0$, the density takes the Wigner semi-circular form for large values of $N$. As mentioned in the previous section, the system's activity can be tuned by varying both the speed $v_0$ and the tumbling rate $\gamma$. Initially, we adjust the activity of the system by varying the parameter $a \in (0,1]$ where, $a$ is defined as 
\beq \label{def_a}
v_0=a N \;,
\eeq
while keeping the tumbling rate fixed at unity. Later, we will also present results due to the variation of the tumbling rate $\gamma$. Here, we will demonstrate that even for the active ISM, the density converges to the Wigner semi-circle in a significant range of parameters. To obtain the density in the steady state, we first evolve the system for a very long time following Eq.~\eqref{eqn_of_motion}. After that, the density profile is obtained simply as the histogram of the positions averaged over many independent realisations. To present density profiles for various system sizes efficiently, we find it useful to show the scaled profile $\tilde{\rho}(z)$ defined by
\beq
\tilde{\rho}(z)=\sqrt{N} \rho(\sqrt{N} z),
\label{eq:scaled_density}
\eeq
where both $\tilde{\rho}$ and $z$ are of $\mathcal{O}(1)$. For further discussion, it is useful to identify three parameter regimes depending on the value of $v_0$. 

\subsection{Weakly active regime}
At zero activity ($v_0=0$), the equilibrium configuration $\{ x^{\rm (eq)}_i\}$ is obtained by equating $dx_i/dt=0$ in Eq.~\eqref{eqn_of_motion} and solving the resulting equations. This implies that the equilibrium configuration $\{x^{\rm (eq)}_i\}$ corresponds to the zeros of Hermite polynomials~\cite{Agarwal_JSP_2019}. For small $v_0$, the fluctuations of the particle's positions around their equilibrium positions are small. This is illustrated in Figure \ref{figure_density_small}(a) where we observe sharp peaks in the density about the equilibrium positions. For instance, with just two particles ($
N=2$), the peaks are located (in the unscaled variable) around 
$x=\pm 0.71$ which are the zeros of the Hermite polynomials of degree two. Similarly, for  $N=4$ the locations of the peaks are at $ x=\pm 1.65$ and $x=\pm 0.52$ which again are Hermite zeros of degree four. As long as the width of these peaks remains much smaller than the inter-particle distance, we refer to the system as being in {\it the weakly active regime}. In this regime, the entire system resembles a crystal where particles are ordered around their equilibrium positions, as illustrated in Fig.~\ref{figure_density_small}(a). Therefore, the density, when coarse-grained at a scale larger than the inter-particle distance, is governed in this regime by the Wigner semi-circle in the large-$N$ limit with finite support inside $[-x_e,x_e]$ where $x_e = \sqrt{2 N}$ (see Fig.~\ref{figure_density_large}(a)). 

\begin{figure}[t]
\includegraphics[width=1.0\linewidth]{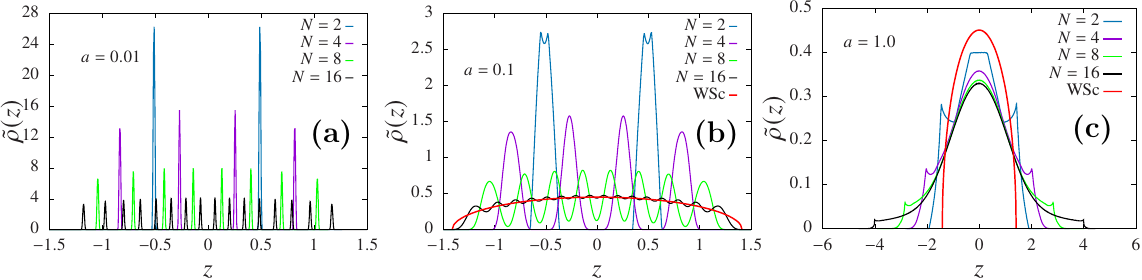}
\caption{The total (scaled) density profiles $\tilde{\rho}(z)$ for different (small) system sizes with several values of the activity, $v_0 = a N$ with $a=0.01$ in (a), $0.1$ in (b) and $1.0$ in (c)  show the crossover: from sharply peaked density profile in (a) at very small activity to bell shaped density profile in (c) at very large activity via an intermediary WSc profile in (b). For all the panels, the tumbling rate $\gamma$ is kept fixed at $1$.}
\label{figure_density_small}
\end{figure}

\begin{figure}[t]
\includegraphics[width=1.0\linewidth]{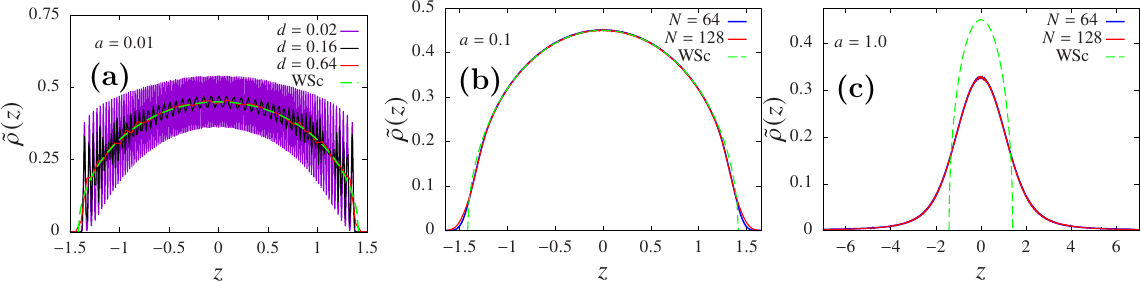}
\caption{The transition in the (scaled) density profile $\tilde{\rho}(z)$ is shown once again with the same set of parameters as in Fig.~\ref{figure_density_small}, but now for large system sizes. Except for panel (a), we display density profiles for two different system sizes, $N=64$ and $N=128$. In panel (a), the density profile is presented for a system size of $N=128$ with an activity parameter of $a=0.01$, where the peaks in the profile smoothen out as the bin size 
$\frac{d}{\sqrt{N}}$ increases, approaching the Wigner semi-circle (denoted as WSc). In the middle panel (b) with an activity parameter of $a=0.1$, the density profile in the bulk approximately matches the WSc form but deviates at the edges. The deviation from the WSc profile increases with system size. The form as well as the support of the density profile has changed significantly in (c) with $a=1.0$. The density forms a bell-shaped profile extending approximately to $[-aN, aN]$ (depicted separately in Fig.~\ref{figure_edge_scaling}). However, it is important to note that the $\sqrt{N}$ scaling in the density profile remains consistent throughout the bulk. For panels (b) and (c), a bin size of $0.01/\sqrt{N}$ is used to construct the histograms of density profiles. The details of the numerical simulation are given in \ref{numerical_details}.}
\label{figure_density_large}
\end{figure}

\subsection{Intermediate activity regime}
As the activity increases, particle fluctuations around their equilibrium positions grow, becoming comparable to the distance to their nearest neighbors. Consequently, the density profile becomes smoother. For example in Fig.~\ref{figure_density_small}(b) we show data for a system with $N=16$ particles where we see that the peaks in the steady-state density profile almost vanish, creating a density profile that aligns well with the Wigner semi-circle form. In Fig.~\ref{figure_density_large}(b), for the activity parameter $a = 0.1$, the steady-state density profiles for two different system sizes match extremely well with the WSc form, except at the edges. We refer to this as {\it the intermediate activity regime}, where the steady-state density profile loses the peaks, at least in the bulk, and is well described by the WSc form even at finite $N$. The quantitative distinction between the weakly active and intermediate activity regimes is achieved by computing the Lindemann's ratio (shown below). Thus, as observed in the case of the active DBM~\cite{Touzo_EPL_2023}, the density profiles in the active ISM also maintain their Wigner semi-circular form in the bulk even at significantly high levels of activity. Notice that near the edges, the density profiles exhibit slight deviations from the Wigner semi-circular form. However, it is important to observe that, here, there are no ``wings", a clear distinction from the active DBM discussed in Refs.\cite{Touzo_EPL_2023,touzo_arxiv_2023}. This difference 
is related to the distinctive behavior of the edge particles compared to those in the bulk of the active DBM~\cite{touzo_arxiv_2023} (see Sec. \ref{sec:var_posn_largeN} near Eq.~\eqref{eqn:mathcal_vb} below), as we now discuss in detail. Consider the equation of motion for one of the rightmost particles (with label $i$ close to $N$)
\beq
\frac{dx_i}{dt}=- x_i+\sum_{\substack{j=1\\j \neq i}}^{N} \frac{\kappa}{(x_i-x_j)^\alpha}+ v_0 \sigma_i(t)  \quad (\alpha,\kappa) = \begin{cases} (1,1) \text{ for the active DBM} \\ (3,2) \text{ for the active ISM} \end{cases} \;.
\eeq
In the intermediate activity regime that we consider in this section, the average position of the particle is obtained by balancing the interaction term with the harmonic force. Let us thus focus on the interaction term. If we assume that all the particles contribute to the same order in the total interaction force felt by particle $i$, we can write 
\beq \label{xeq_edge}
\sum_{\substack{j=1\\j \neq i}}^{N} \frac{\kappa}{(x_i-x_j)^\alpha} \simeq N \kappa \int_{-\infty}^{+\infty} dx \frac{\rho_{sc}(x)}{(x_i-x)^\alpha} = N^{1-\alpha/2} \frac{\kappa}{\pi} \int_{-\sqrt{2}}^{\sqrt{2}} dx' \frac{\sqrt{2-x'^2}}{(x_i/\sqrt{N}-x'^2)^\alpha} \;,
\eeq
where in the last step we have made a change of variable $x=\sqrt{N}x'$. Since $x_i/\sqrt{N}\sim \sqrt{2}$, we see that the total interaction term is of order $\sqrt{N}$ for the active DBM, which when balanced with the harmonic force leads to $x_i^{\rm (eq)} \sim \sqrt{N}$, which is the correct behavior. On the other hand, for the active ISM, the same argument leads to $x_i^{\rm (eq)} \sim N^{-1/2}$, while we also observe $\sqrt{N}$ in this case. This discrepancy is due to the fact that, for the active IS, the total force is actually dominated by the few particles which are closest to particle $x_i$. Indeed, since the equilibrium positions $x_i^{\rm (eq)}$ of the particles in both models are given by the zeros of Hermite polynomials, we can use the asymptotic behavior of these zeros at the edge (see e.g. Ref. \cite{Hermite_zeros_edge} or \cite{Agarwal_JSP_2019}) to write for $m=N-i=\mathcal{O}(1)$,
\beq
x_{N-m}^{\rm (eq)} = \sqrt{2N} + 2^{-1/3} (2N)^{-1/6} a_m + \mathcal{O}(N^{-1/2}) \;,
\eeq
where $a_m <0$ is the $m^{\rm th}$ zero of the Airy function. This implies that the typical spacing between the edge particles in both models is of order $N^{-1/6}$. Thus, the force exerted on an edge particle $i$ by its neighbors is of order $N^{1/6}$ in the active DBM, which is negligible compared to the total interaction force. By contrast, in the active ISM, it is of order $\sqrt{N}$, which is the order required to balance the harmonic force. This explains why the distance between particles, and thus the positions of edge particles cannot fluctuate much in the active ISM. By contrast in the active DBM, the edge particles can fluctuate much more, leading to the wings observed in Ref.\cite{Touzo_EPL_2023}.

\subsection{Strongly active regime}

When $v_0$ becomes significantly large, particles at the edge of the system can explore a much broader range within the harmonic trap compared to the passive case. In this regime, the particle density undergoes a marked change from the typical semi-circular profile observed in passive systems. Instead, it adopts a bell-shaped form, reflecting the enhanced movement and redistribution of particles due to the increased activity [see Fig.~\ref{figure_density_small}(c) and ~\ref{figure_density_large}(c)]. Remarkably, as the density profile shifts from a Wigner semi-circle to a bell-shaped distribution, the scaling behavior with the system size as in Eq.~\eqref{eq:scaled_density} persists. The bulk of the density is thus still contained over an interval of size $\sim \sqrt{N}$. However, the total support of the density profile undergoes an approximate shift to $[-aN,aN]$, contrasting with the previous support of $[-\sqrt{2N},\sqrt{2N}]$ (see Fig.~\ref{figure_edge_scaling}). Additionally, near the edges, the density profile exhibits a power-law decay $\sim 1/x^3$ (see Fig.~\ref{figure_edge_scaling}). The $1/x^3$ power-law decay is not limited to the active ISM; it has been observed in various other systems. For instance, it also arises in the active DBM under extreme activity conditions ($v_0 \sim \mathcal{O}(N)$), as shown in Ref.\cite{Touzo_EPL_2023}. Note that the rescaled velocity is defined as $v_0=\frac{\bar{v}_0}{\lambda^{3/4} g^{1/4}}$, so that this strongly active regime can be reached either by increasing $v_0$ or equivalently by decreasing the interaction strength $g$. This is consistent with the observation in Ref.\cite{Touzo_EPL_2023} that this behavior corresponds to the limit where the interaction effectively behaves as a contact interaction. The $1/x^3$ decay seems to be universal, independent of the specific interaction potential, provided it involves infinite repulsion upon contact. Nevertheless, our current understanding of this decay still lacks theoretical insights.

Note that this tail seems to contain a number of particles which is $\mathcal{O}(1)$. Indeed, the scaling law $N^2\rho(x) \sim (N/x)^3$ observed in Fig.~\ref{figure_edge_scaling} implies $\rho(x) \sim N/x^3$, so that the number of particles in the tail is of the order of
$\int_{\sqrt{N}}^N dx \, \rho(x) \sim N \int_{\sqrt{N}}^N \frac{dx}{x^3} \sim \mathcal{O}(1) $. Thus, this $1/x^3$ tail exhibits strong sample-to-sample fluctuations.

\begin{figure}[t]
\includegraphics[width=1.0\linewidth]{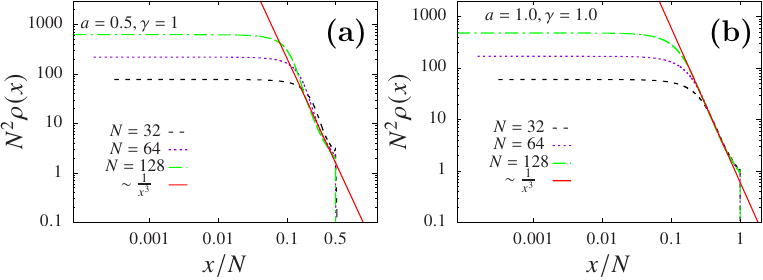}
\caption{Showing the total support of the density profile $\rho(x)$ approximates to $[-a N,aN]$ in the large activity limit. The steady-state density profiles are drawn as a function of $x/N$ for three different system sizes which seem to converge and show a sharp drop at $x=aN$. Near the edges, the density profiles (dashed lines) decay as $\sim 1/x^3$ (red solid line). }
\label{figure_edge_scaling}
\end{figure}

Recall that the system's activity can also be enhanced by reducing the tumbling rate $\gamma$. In Fig.~\ref{figure_varying_gamma}, we demonstrate a crossover in a system of $N=64$ particles, transitioning from a sharply peaked to a smooth Wigner semi-circle density profile (at least near the center of the trap) as $\gamma$ is reduced, with the activity strength $v_0$ held constant at a very small value. With such a small $v_0$, the strongly active regime cannot be achieved solely by reducing $\gamma$ as shown in the `phase' diagram (see Fig.~\ref{fig:phase_diagram_together}). Thus, the system's activity is mainly controlled by $v_0$ rather than $\gamma$.

In the strongly active region, where the density profile deviates significantly from the Wigner semi-circle and attains a bell-shaped profile, it has a sharp cut-off at $\pm v_0$ when $\gamma$ remains sufficiently large (see Fig.~\ref{figure_edge_scaling}). For smaller $\gamma$, two slightly broadened peaks appear near the edges ($\pm v_0$), creating a wing-like structure as shown in Fig.~\ref{figure_edge_wings}.}
Indeed in this regime, the argument used above to explain the absence of wings for the active ISM is not valid anymore, since the shape of the density is not anymore primarily determined by the balance between the interactions and the confining potential. The position of the wings can be roughly obtained by balancing, for the edge particles, the driving velocity term with $\sigma_i=+1$ and the harmonic term, leading to $x_i\sim v_0 \sim N$ (while in the intermediate regime of the active DBM, the position of the wings is $\sim \sqrt{N}$). Note that these wings are only observed when $\gamma$ is small enough ($\gamma \lesssim 0.1$), since it requires that the particles spend a lot of time near the fixed points for which $\sigma_i=+1$ for the edge particles.

\begin{figure}[t]
\includegraphics[width=1.0\linewidth]{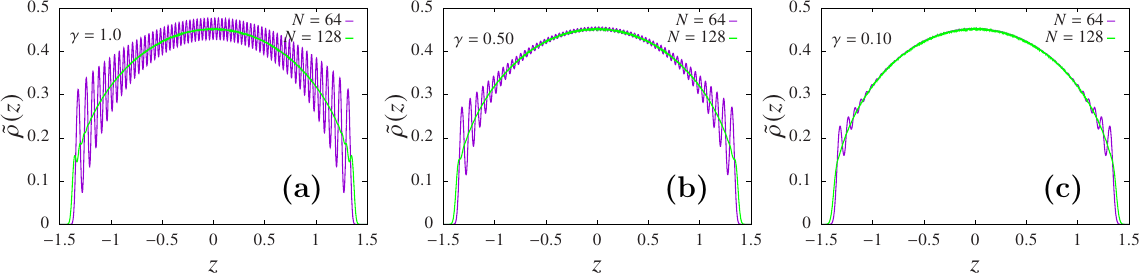}
\caption{Plot of the density $\bar{\rho}(z)$ vs $z$ showing the melting of the `crystal' to a `liquid' in a system of $64$ particles because of reducing the tumbling rate $\gamma$ keeping the activity parameter $a$ fixed at $a=0.025$. For the system size $N=128$ with $a=0.025$, the system has already crossed over to the liquid state characterized by a smooth density profile.}
\label{figure_varying_gamma}
\end{figure}

\begin{figure}[h]
\includegraphics[width=1.0\linewidth]{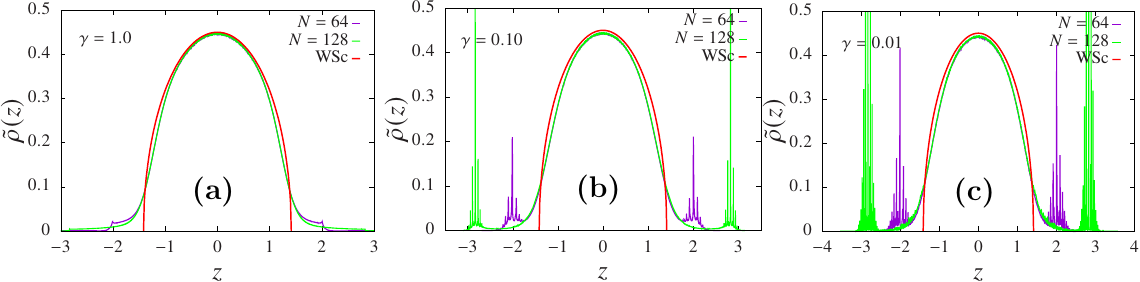}
\caption{Demonstration of the emergence of wing-like structures in the density profile within the strongly active region ($a=0.25$) as a result of the reduced tumbling rate, $\gamma$.}
\label{figure_edge_wings}
\end{figure}

\subsection{Discussion of crossovers}
\begin{figure}[t]
\includegraphics[width=1.0\linewidth]{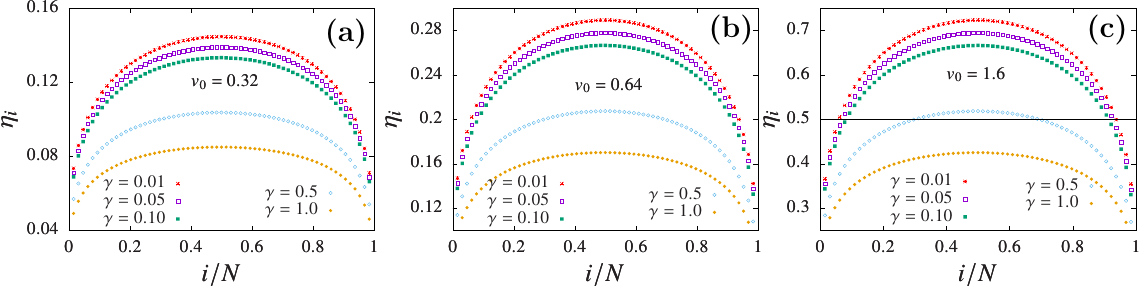}
\caption{Numerically computed Lindemann's ratio $\eta_i$ (points) as a function of $i/N$ for different values of the tumbling rate $\gamma$ within a system comprising $N=64$ particles. Three panels correspond to three different values of the speed $v_0$, $0.32, 0.64$ and $1.6$ in (a),(b) and (c) respectively. Additionally, we have drawn a solid black (horizontal) line at $0.5$ indicating the cases where the standard deviation of particle positions becomes exactly equal to $50\%$ of the average distance to nearest neighbour. For a quantitative distinction between crystal and liquid regimes, we use this as a reference line above which we consider the system to be in the liquid regime and below which in the crystal regime. In the first two panels, the system remains in the crystal regime regardless of the $\gamma$ values analyzed. However, in the third panel with speed $v_0=1.6$, the system stays in the crystal regime for higher values of $\gamma$ ($\gamma \gtrsim 1$). Conversely, for smaller $\gamma$ values ($\lesssim 0.5 $), the Lindemann ratio exceeds $0.5$ at the center, indicating a liquid phase.}
\label{figure_linemanns_ratio}
\end{figure}

\begin{figure}[h]
\includegraphics[width=1.0\linewidth]{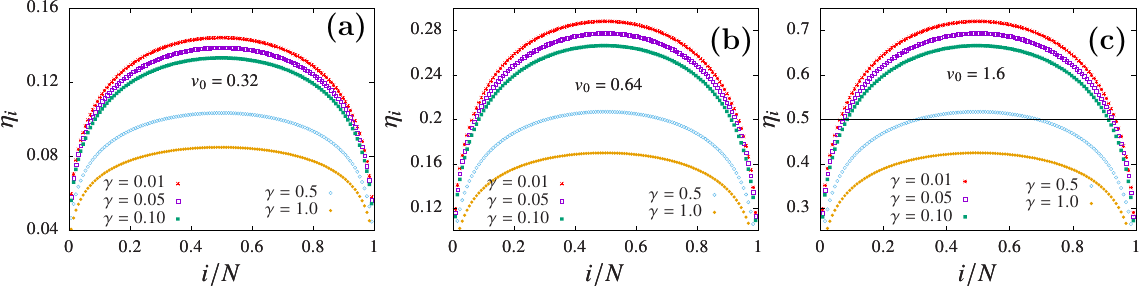}
\caption{Lindemann's ratio $\eta_i$ as a function of $i/N$ for different values of the tumbling rate $\gamma$ is shown in Fig.~\ref{figure_linemanns_ratio}, now within a system comprising $N=128$ particles. Comparing these figures with that in Fig.~\ref{figure_linemanns_ratio}, it is now evident that the quantitative behavior of the system regarding the crystal to liquid crossover remains unchanged with changing system size, provided the speed $v_0$ is $\mathcal{O}(1)$ (independent of $N$).}
\label{figure_linemanns_ratio_128}
\end{figure}

Thus, based on values of the parameters $v_0$ and $\gamma$, the steady-state density profile has three different forms. When the activity is extremely low, the particles stay close to their equilibrium positions and fluctuate within a small region. This leads to a density profile with sharp peaks, exhibiting a crystal-like phase. As the activity increases, particle fluctuations around their equilibrium positions grow, becoming comparable to the distance to their nearest neighbors. The density profile then becomes smoother, resembling a liquid phase. To quantify this crossover further, we numerically compute the  Lindemann's ratio defined in Eq.~\eqref{eq:lindemanns_ratio}. This ratio compares the standard deviation of a particle's position to the average gap between the particle and its nearest neighbour. The numerical values of this ratio are shown in Fig.~\ref{figure_linemanns_ratio} for different values of the activity parameter $v_0$ and the tumbling rate $\gamma$ in a system with a total $64$ particles. Notably, we introduce a reference line at $0.50$, representing scenarios where particle fluctuations are exactly $50\%$ of the average gap. When Lindemann's ratio at the center exceeds this reference value, we consider the system to be in a `liquid phase'; when it is below, it is in a crystalline phase. It's evident that a system comprising $64$ particles with speed $v_0=0.32$ and $0.64$ consistently remains within the crystalline phase, as illustrated in Fig.~\ref{figure_linemanns_ratio}(a) and (b). Conversely, for $v_0=1.6$, the system resides in the crystalline phase only when the tumbling rate $\gamma$ is greater than $0.5$; below this value, it is in the `liquid phase'. The corresponding scenarios are illustrated in Fig.~\ref{figure_linemanns_ratio_128} for a system comprising of $128$ particles. From Figs.~\ref{figure_linemanns_ratio} and \ref{figure_linemanns_ratio_128}, it is now clear that the `crystal' to `liquid' transition occurs at $v_0 \sim \mathcal{O}(1)$, independent of system size. 


\begin{figure}[t]
\includegraphics[width=0.8\linewidth]{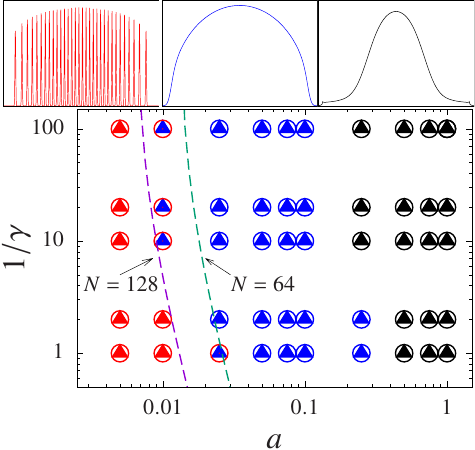}
\caption{Lower diagram depicts three distinct regions in the active ISM within the parameter space represented by $a$ and $1/\gamma$, with red, blue, and black points representing weakly active, intermediate, and strongly active regimes, respectively. Red points signify scenarios where Lindemann's ratio $\eta_i$ at the center of the trap ($i=N/2$) is less than $0.5$ (see Fig.~\ref{figure_linemanns_ratio_128}). Blue points represent intermediate active regions where the Lindemann's ratio at the trap center exceeds $0.5$, but the values of $\chi$ defined in Eq. \eqref{eq:def_chi} remain below $0.1$ (see Table.~\ref{tab:my-table}). The strongly active regime (indicated by black points) corresponds to cases where the value of $\chi$ exceeds $0.1$. Density profiles in the corresponding regimes are shown schematically in the top panels. Two different system sizes are represented:  circles ($\circ$) correspond to $N=64$, while triangles ($\blacktriangle$) represent $N=128$. Notice that the second transition (from blue to black) occurs at $a \sim \mathcal{O}(1)$, which is independent of $N$. In contrast, the first transition (from red to blue) occurs at $a$ values that depend on $N$, as depicted by two schematic (dashed) lines for the different system sizes.} 
\label{fig:phase_diagram_together}
\end{figure}

For the second crossover, from a smooth Wigner semi-circular density profile to a bell-shaped density profile, we numerically compute the values of the distance measure $\chi$ as defined in Eq.~\eqref{eq:def_chi} across various combinations of parameters $a$ and $\gamma$. Table \ref{tab:my-table} presents the values of $\chi$ for two distinct system sizes, $N=64$ and $128$. Whenever the value of $\chi$ exceeds $0.1$, we refer to this as {\it the strongly active regime}. Based on the values of the Lindemann's ratio and the quantity $\chi$, we draw a `phase' diagram in the $a$, $1/\gamma$ plane shown in Fig.~\ref{fig:phase_diagram_together} separating three distinct phases which correspond to the weakly active, the intermediate active and the strongly active regions.

In summary, we numerically compute the steady-state density profile in three activity regimes: weakly active, intermediate active and strongly active regime. In these regimes, the density profile exhibits distinct characteristics: in the weakly active regime, it features sharp peaks akin to a crystalline phase, which vanish (at least in the bulk) in the intermediate active region, resembling a transition to a liquid-like phase. In the strongly active region, the density profile assumes a bell-shaped form. To distinguish these three regions more quantitatively, we look at the Lindemann's ratio $\eta_i$ (defined in Eq.~\eqref{eq:lindemanns_ratio}) and the quantity $\chi$ (defined in Eq.~\eqref{eq:def_chi}). Based on their numerical values, we made a `phase' diagram in the $a$, $1/\gamma$ plane. In the following, we discuss our theoretical understanding of the fluctuations in particle positions using Hessian calculations.

\section{Variance and covariance of the particle displacements}
\label{section_hessian}
In this section, we compute the correlation between the displacements of different particles in the small noise limit $(v_0 \to 0)$. To this end, we employ the Hessian-based technique developed in Ref.~\cite{Agarwal_JSP_2019}. A striking feature about the model in Eq.~\eqref{eqn_of_motion} is that for $v_0=0$, the positions of the particles $\bold{x}^{\rm{(eq)}} = (x_1^{(\rm{eq})}, x_2^{(\rm{eq})},...,x_N^{(\rm{eq})})$ are given by the zeroes of the Hermite polynomial of degree $N$, \emph{i.e.,} $H_N\left(x_i^{(\rm{eq})}\right) = 0$. Moreover, for $N \to \infty$, the equilibrium particle density converges to the Wigner semi-circle which has a finite support on $[-x_e,x_e]$ where $x_e=\sqrt{2N}$. For small values of $v_0$, we expect the position $x_i$ of the $i$-th particle to deviate only slightly from its equilibrium value $x_i^{(\rm{eq})}$. Let us denote this displacement by $\delta x_i$ and write
\beq
x_i=x^{\rm{(eq)}}_i+ \delta x_i.
\eeq
Assuming that the typical fluctuations of the particle positions $\delta x_{i}$ are small compared to the typical size of the gaps $x_{i+1}^{(\rm{eq})}-x_{i}^{(\rm{eq})} \sim 1/\sqrt{N}$, one can linearise the equation of motion in \eqref{eqn_of_motion} in $\delta x_i$ to obtain
\begin{align}
\frac{d \left( \bold{\delta x} \right)}{dt} \simeq - \bold{H}_{\text{ISM}} \bold{\delta x} + v_0 \boldsymbol{\sigma}, \label{linear-eqn-mot}
\end{align}
with $\bold{\delta x} = \left( \delta x_1, \delta x_2,...,\delta x_N \right)^{T}$ and $\boldsymbol{\sigma} = \left( \sigma _1(t), \sigma _2(t),..,\sigma_N(t) \right)^{T}$. Moreover, $\bold{H}_{\text{ISM}}$ denotes the Hessian matrix for the ISM and is given completely in terms of the zeroes of the Hermite polynomial as~\cite{Agarwal_JSP_2019}
\begin{equation}
\left( \bold{H}_{\text{ISM}}\right) _{ij} = \delta _{i,j} \left[ 1+\sum_{\substack{k=1\\k \neq i}}^{N}  \frac{6}{\left( x_{i}^{(\rm{eq})}-x_{k}^{(\rm{eq})} \right)^4}\right] - \frac{6 \left(1-\delta _{i,j} \right) }{\left( x_{i}^{(\rm{eq})}-x_{j}^{(\rm{eq})} \right)^4}. \label{Hessian-CM}
\end{equation}
Although we have linearized the equation of motion, solving Eq.~\eqref{linear-eqn-mot} analytically still turns out to be difficult due to the dynamics of the $\sigma$-variables. Therefore, in order to simplify the dynamics further, we will focus on the $\gamma \to 0^+$ limit, which implies that the $\sigma$-variables remain fixed to their initial values throughout the time evolution. However, these initial values still fluctuate for different realizations. Now, for a given realisation of $\boldsymbol{\sigma}$, we anticipate our system to go to a unique fixed point at large times. For the case of the active Dyson Brownian motion, this was seen in Ref.~\cite{touzo_arxiv_2023} using extensive numerical simulations. Anticipating this to be true for the active ISM also, we can then write the steady-state displacements from Eq.~\eqref{linear-eqn-mot} as
\begin{align}
 \bold{\delta x} \simeq  v_0 ~\bold{H}_{\text{ISM}}^{-1}~\boldsymbol{\sigma}.
\end{align}  
We now use this solution to calculate the variance and covariance of the displacement variables. To do this, we first recall that $\sigma_i$ is chosen from $\pm 1$ with equal probability independently of the index $i$. Hence, we obtain $\langle \sigma _i \rangle = 0$ and $\langle\sigma _i \sigma _j \rangle = \delta _{i,j}$. This allows us to write the expressions for the mean and the covariance as
\begin{align}
 \langle \delta x_i \rangle &\simeq 0 \\
 \langle \delta x_i \delta x_j \rangle & \simeq v_0 ^2 \left( \textbf{H}_{\text{ISM}}^{-2} \right)_{ij}\;.
\label{covariance_form}
\end{align}
Note that at leading order in $v_0$, the average of $\delta x_i$ vanishes, which means that the average position $\bar x_i$ of the $i^{th}$ particle coincides with its equilibrium position $x_i^{\rm (eq)}$. This implies in particular that $s_i^2=\langle \delta x_i^2 \rangle$ and $g_{i,n}^2=\langle (\delta x_{i+n} -\delta x_i)^2 \rangle$.

\subsection{Inverting the Hessian matrix $\bold{H}_{\text{ISM}}$}
Given the complicated form of the matrix $\bold{H}_{\text{ISM}}$ in Eq.~\eqref{Hessian-CM}, it might seem difficult to invert it analytically. However, we use the fact that the Hessian matrix of the ISM can be expressed in terms of the Hessian matrix $\bold{H}_{\text{L}}$ of the DBM as $\bold{H}_{\text{ISM}} = \bold{H}_{\text{L}}^2 $ where  $\bold{H}_{\text{L}}$ is given by~\cite{Agarwal_JSP_2019}
\begin{align}
\left( \bold{H}_{\text{L}} \right)_{ij}=\delta_{ij} \Big[1+\sum_{k \neq i} \frac{1}{(x_i^{\rm{(eq)}}-x_k^{\rm{(eq)}})^2}\Big]- \frac{(1-\delta_{ij})}{(x_i^{\rm{(eq)}}-x_j^{\rm{(eq)}})^2}.
\end{align}
The covariance in Eq.~\eqref{covariance_form} then becomes
\begin{align}
 \langle \delta x_i \delta x_j \rangle & \simeq v_0 ^2 \left( \textbf{H}_{\text{L}}^{-4} \right)_{ij}.
\label{covariance_form-2}
\end{align}
It turns out that the matrix $\textbf{H}_{\text{L}}$ can be diagonalized exactly using the Hermite polynomials (see Ref.\cite{Ahmed_INCB_1979}). Its eigenvalues are simply the integers from $1$ to $N$, and the normalized eigenvector $\psi_k$ associated to the eigenvalue $k$ has components given by
\beq
(\psi_k)_i=\frac{u_k\left( x_i^{\rm{(eq)}} \right)}{\sqrt{\sum_{l=1}^{N} u_k\left( x_l^{\rm{(eq)}} \right)^2}}\;,~ ~\text{with}~ u_k(x)=2^{k-1} \frac{(N-1)! H_{N-k}(x)}{(N-k)! H_{N-1}(x)} \;.
\eeq
Using the eigenvalue decomposition
\beq
\left( \textbf{H}_{\text{L}}^{-4} \right)_{ij}=\sum_{k=1}^{N} \frac{(\psi_k)_i (\psi_k)_j}{k^4},
\eeq
the covariance in Eq.~\eqref{covariance_form-2} then simplifies to
\beq
\langle \delta x_i \delta x_j \rangle \simeq v^2_0 \sum_{k=1}^{N} \frac{1}{k^4}~\frac{u_k\left( x_i^{\rm{(eq)}} \right) u_k\left( x_j^{\rm{(eq)}} \right)}{\sum_{l=1}^{N} u_k\left( x_l^{\rm{(eq)}} \right)^2} \;.
\label{corr_posn_formula}
\eeq
From this expression, the variance $\langle \delta x_i ^2\rangle $ turns out to be
\beq
\langle \delta x_i ^2\rangle  \simeq v^2_0 \sum_{k=1}^{N} \frac{1}{k^4}~\frac{u_k\left( x_i^{\rm{(eq)}} \right)^2}{\sum_{l=1}^{N} u_k\left( x_l^{\rm{(eq)}} \right)^2}\;.
\label{var_posn_formula}
\eeq

Note that, since $\mathcal{H}_N(-x)=(-1)^N \mathcal{H}_N(x)$, we have the symmetry $ \langle \delta x_i \delta x_j \rangle = \langle \delta x_{N-i+1} \delta x_{N-j+1} \rangle$. These results 
in Eqs. (\ref{corr_posn_formula}) and (\ref{var_posn_formula})
are similar to the ones found for the active DBM, but with a $1/k^4$ factor instead of $1/k^2$~\cite{touzo_arxiv_2023}. 
Incidentally, this results in a faster numerical convergence of the sum over $k$. Later, we will discuss that this implies some interesting differences between the two models. In what follows, we will analyze the results derived in Eqs.~\eqref{corr_posn_formula} and \eqref{var_posn_formula} in the large $N$ limit and obtain some nice scaling relations for the variance and the covariance.

\subsection{Variance and covariance in the large $N$ limit}
\label{sec:var_posn_largeN}
We now make the same approximations as done in Ref.~\cite{touzo_arxiv_2023} in the limit $N \to \infty$. We first note that even though the summation in Eq.~\eqref{corr_posn_formula} runs from $1$ to $N$, the sum gets dominant contribution from the smaller values of $k$. In particular, for $k \ll N$, the function $u_k\left( x_i^{\rm{(eq)}} \right)$ has a simplified expression~\cite{touzo_arxiv_2023}
\beq
u_k\left( x_i^{\rm{(eq)}} \right) \simeq (2N)^{\frac{k-1}{2}} U_{k-1} \left(\frac{ x_i^{\rm{(eq)}}}{\sqrt{2N}} \right), \label{PS-eq-1}
\eeq
in terms of the Chebyshev polynomial of the second kind $U_k(w)$ of order $k$. Moreover for $|w|<1$, $U_k(w)$ can be written as
\begin{align}
U_k(w)= \frac{\sin\big((k+1) \arccos w \big)}{\sqrt{1-w^2}}\;. \label{PS-eq-2}
\end{align}

\begin{figure}[t]
\includegraphics[width=1.0\linewidth]{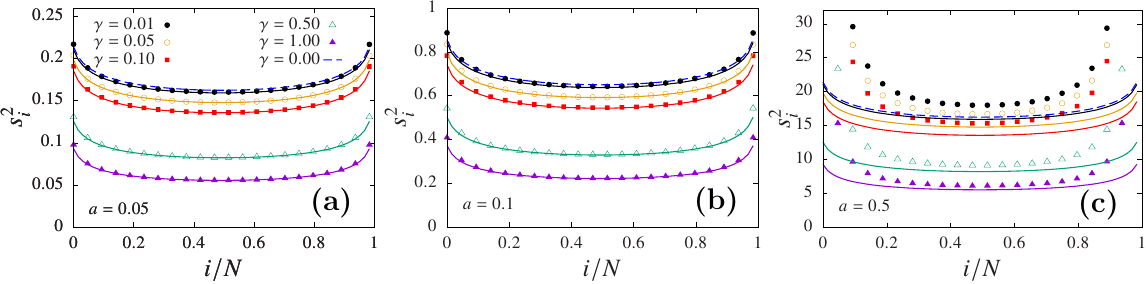}
\caption{Numerically computed variance of particle positions $s^2_i \equiv \langle x^2_i \rangle -\langle x_i \rangle^2$ (points) in the steady-state of active ISM for different values of the tumbling rate $\gamma$ keeping the activity parameter $a$ fixed at $0.05$ in (a), $0.1$ in (b) and $0.5$ in (c) for the system size $N=64$. The dashed blue line represents our theoretical expectation for $\gamma=0^+$ given in Eq.~\eqref{eqn:mathcal_vb}. The solid lines show our prediction for finite $\gamma$ given in \eqref{vargamma_largeN}. In the first two panels, we observe a very good matching between numerical and analytical results for any value of the tumbling rate $\gamma$. However, in the last panel where the activity is very high, the approximation used in the theoretical calculation is not valid anymore. Indeed we observe that there is no more agreement with the numerics in this case.}
\label{figure_var_posn}
\end{figure}

To fully simplify the correlation in Eq.~\eqref{corr_posn_formula}, we still need to simplify the denominator $\sum_{l=1}^{N} U_{k-1}\left( \frac{x_l^{\rm{(eq)}}}{\sqrt{2 N}} \right)^2$ in this expression. Remember that as $N$ becomes large, the density of the Hermite zeroes approaches the Wigner semi-circle. Moreover, the difference between successive roots $\left[ \frac{x_l^{\rm{(eq)}}}{\sqrt{2 N}} - \frac{x_{l-1}^{\rm{(eq)}}}{\sqrt{2 N}} \right] \sim \frac{1}{N}$. Therefore, we can change the aforementioned summation to an integration with an appropriate measure, namely
\begin{align}
\sum_{l=1}^{N} U_{k-1}\left( \frac{x_l^{\rm{(eq)}}}{\sqrt{2 N}} \right)^2 \simeq N \int_{-1}^{1} dw \frac{2 \sqrt{1-w^2}}{\pi} U_{k-1}(w)^2 = N \;. \label{PS-eq-3}
\end{align}
Finally plugging the approximations from Eqs.~\eqref{PS-eq-1}-\eqref{PS-eq-3} in Eq.~\eqref{corr_posn_formula} and taking the $N \to \infty$ limit, we find that the covariance $\langle \delta x_i \delta x_j \rangle$ satisfies the scaling relation
\begin{align}
\langle \delta x_i \delta x_j \rangle  \simeq \frac{{v}^2_0}{N}~ \mathcal{C}\left( \frac{x^{\rm{(eq)}}_i}{\sqrt{2N}},\frac{x^{\rm{(eq)}}_j}{\sqrt{2N}} \right)\;,  \label{PS-eq-4}
\end{align}
with the scaling function $\mathcal{C}(w,z)$ defined as
\begin{align}
\mathcal{C}(w,z) & =\sum_{k=1}^{N} \frac{1}{k^4} U_{k-1}(w) U_{k-1}(z),  \label{PS-eq-5}
\end{align}
where both the scaled variables $w$ and $z$ lie between $[-1,1]$. In fact, the sum over $k$ in (\ref{PS-eq-5}) can actually be computed explicitly, leading to
\begin{align}
\mathcal{C}(w,z)& =\frac{c  (\arccos w,\arccos z ) }{\sqrt{1-w^2} \sqrt{1-z^2}},~~\text{with }  \label{eqn:mathcal_Cb}\\
c(u,v)&=\frac{1}{6} \Big[\frac{uv}{2} (u^2+v^2) +\frac{\pi}{4} (|u-v|^3-(u+v)^3+\pi^2 uv \Big],  \label{PS-eq-7} \\
& = \frac{v}{12}(\pi-u)(2 \pi u -u^2-v^2)~\text{for}~ u>v \;.
\end{align}
From the expression of the covariance, one can easily get the variance of the position for a single particle as
\beq
\langle \delta x^2_i \rangle \simeq \frac{v^2_0}{ N}~ \mathcal{V}\left(\frac{x^{\rm{(eq)}}_i}{\sqrt{2N}}\right),~\text{with}~ \mathcal{V}(w)=\frac{\arccos^2 w(\pi -\arccos w)^2}{6(1-w^2)}.  \label{eqn:mathcal_vb}
\eeq
In Fig.~\ref{figure_var_posn}, we compare this analytical formula in Eq.~\eqref{eqn:mathcal_vb} with the numerically computed variance of particle positions in the steady state. The comparison is done across various values of the activity parameters $v_0$ and $\gamma$ in a system with $N=64$. The formula provided in Eq.~\eqref{eqn:mathcal_vb} (labeled by the dashed line in Fig.~\ref{figure_var_posn}) appears to be valid for sufficiently large values of $v_0$. However, it is evident that the tumbling rate $\gamma$ needs to be very low.

Since for a general particle $x_i^{\rm(eq)} \sim \sqrt{N}$, the only real $N$ dependence in~\eqref{PS-eq-4} and~\eqref{eqn:mathcal_vb} is in the prefactor and we find that the variance and covariance of the particle positions decay as $\langle \delta x_i^2 \rangle \sim 1/N$ for large $N$. This is the case even for particles which are deep inside the bulk, i.e. when $|x_i^{\rm(eq)}| \ll \sqrt{N}$, since $\mathcal{V}(w)$ is finite near $w=0$ ($\mathcal{V}(w)=\frac{\pi^4}{96}+O(w^2)$). This is similar to the behavior obtained in Ref.~\cite{touzo_arxiv_2023} for the active Brownian motion. However, the main difference is that, due to the faster convergence of the series in Eq.~\eqref{PS-eq-5}, these results are also valid for the edge particles. Indeed, expanding $\mathcal{V}(w)$ near the edge (since $x_i^{\rm(eq)} \sim \sqrt{2N}$ for the edge particles, we replace $w = (1-\epsilon)$ in Eq.~\eqref{eqn:mathcal_vb} and take the $\epsilon \to 0^+$ limit) we find
\beq
\mathcal{V}(1-\epsilon) = \frac{\pi^2}{6} -\frac{\pi}{3} \sqrt{2\epsilon} +\mathcal{O}(\epsilon).
\label{approximate_vb}
\eeq
We see that there is a finite limit as $\epsilon \to 0^+$, which indicates that for large $N$ the variance decays as $\langle \delta x^2_i \rangle \sim 1/N$ even for the edge particles. This is very different from the active DBM which exhibits different scaling behaviors for the bulk and the edge particles~\cite{touzo_arxiv_2023}. For the active DBM, the equivalent of the function $\mathcal{V}(w)$ diverges near the edges, which shows that the $1/N$ scaling breaks down, and an alternative analysis is required to obtain the variance of the position of the edge particles. Nothing of the sort seems to occur for the active ISM: the leading order decay of the variance is the same regardless of whether the particle is in the bulk or at the edge. All of this is also true for the covariance \eqref{PS-eq-4}.

As a remark, one can use \eqref{approximate_vb} along with the asymptotics of the Hermite zeros at the edge \eqref{xeq_edge} to obtain the behavior of the edge particles, for which $m=N-i=\mathcal{O}(1)$,
\beq
\langle \delta x^2_{N-m} \rangle \simeq \frac{v^2_0}{ N} \left(\frac{\pi^2}{6}- \frac{\pi}{3} \sqrt{-a_m} \, N^{-1/3}+\mathcal{O}(N^{-2/3}) \right),~~~\text{(for edge particles)}, \label{PS-eq-10}
\eeq
where $a_m$ is the $m^{th}$ zero of the Airy function.

Comparing the typical amplitude of position fluctuations to other existing length scales in the system, one can derive the criterion of the crossovers observed in the steady-state density profile. We first compare the typical displacements $(\sim v_0 / \sqrt{N})$ of the particle  with the typical inter-particle separation $(\sim 1/ \sqrt{N})$ and we get Lindemann's ratio
\begin{align}
\eta_i =\frac{\sqrt{\langle  \delta x^2_i \rangle}}{\left (x^{\rm (eq)}_{i+1}-x^{\rm (eq)}_i \right ) } \sim \frac{v_0 ~\sqrt{N}}{\sqrt{N}}  \sim v_0.
\end{align}
Thus, if $v_0$ is very small ($v_0 \ll 1$), the particle fluctuations are much smaller than the mean spacing between successive particles. This situation corresponds to a density profile with sharp peaks, which we refer to as the weakly active regime. On the other hand, comparing the amplitude of the fluctuations with the size of the support of the semi-circle $x_{e} = \sqrt{2N}$ gives the ratio
\begin{align}
\frac{\sqrt{\langle \delta x _i^2 \rangle}}{x_{e} } \sim \frac{v_0}{ N} \;.
\end{align}
This suggests that when $v_0 \sim \mathcal{O}(N)$, the particle fluctuations are of the order of the total support, which indicates a crossover to the strongly active regime. 
Thus the transition from weakly active to intermediate active region occurs when $v_0 \sim \mathcal{O}(1)$ and the second transition, from intermediate to strongly active region happens at $v_0 \sim \mathcal{O}(N)$. Notice that the transition regions are consistent with the `phase' diagram shown in Fig.~\ref{fig:phase_diagram_together} as well. This argument also explains why in the strongly active regime, while the total support of the density is $[-aN,aN]$, the bulk of the density is still contained in an interval of size $\sim \sqrt{N}$. Indeed in this regime, the density is dominated by the fluctuations of the particles, which are typically of order $\sqrt{\langle\delta x_i^2 \rangle} \sim v_0/\sqrt{N} = a \sqrt{N}$ (assuming that the \eqref{eqn:mathcal_vb} still gives the correct order of magnitude for the typical fluctuations in this regime).
\\

\noindent {\bf Extension to finite $\gamma$.}
It is actually possible to extend the present results to finite values of $\gamma$. The derivation was presented in a recent paper \cite{Riesz2024}. The idea is to start again from the linearized equation of motion \eqref{linear-eqn-mot} and to solve it in Fourier space to obtain an expression of the covariance $\langle\delta x_i \delta x_j\rangle$ for any value of $\gamma$. Going back to real space and taking the large $N$ limit, we arrive at
\begin{eqnarray} \label{covgamma_largeN}
&&\langle \delta x_i \delta x_j \rangle \simeq \frac{v_0^2}{N} \mathcal{C}^{\gamma}\left( \frac{x^{\rm (eq)}_i}{\sqrt{2N}}, \frac{x^{\rm (eq)}_j}{\sqrt{2N}} \right) \quad , \\ &&\mathcal{C}^{\gamma}(u,v) = \sum_{k=1}^\infty \frac{1}{k^2(k^2+2\gamma)} \frac{\sin(k \arccos u)}{\sqrt{1-u^2}} \frac{\sin(k \arccos v)}{\sqrt{1-v^2}} \;.
\end{eqnarray}
For the variance this reads
\begin{equation} \label{vargamma_largeN}
\langle \delta x_i^2 \rangle \simeq \frac{v_0^2}{N} \mathcal{V}^{\gamma}\left( \frac{x^{\rm (eq)}_i}{\sqrt{2N}} \right) \quad , \quad \mathcal{V}^\gamma(u) = \sum_{k=1}^\infty \frac{1}{k^2(k^2+2\gamma)} \frac{\sin^2(k \arccos u)}{1-u^2}\;.
\end{equation}
In the limit $\gamma \to 0$ we recover the previous results \eqref{PS-eq-4} and \eqref{eqn:mathcal_vb}, while for $\gamma \gg 1$ we recover the results for the passive inverse-square power-law model derived in Ref.\cite{touzo_arxiv_2023}, with an effective temperature $v_0^2/(2\gamma)$. Here however the sums cannot be computed explicitly. In Fig.~\ref{figure_var_posn}, we see that the agreement with numerical simulations is very good for small values of $a$ (up to $a\sim 0.1$), for all values of $\gamma$.
Note that the variance is a decreasing function of $\gamma$. Thus as $\gamma$ increases, keeping $v_0$ fixed, the particles become more localized, which explains the observation in Fig.~\ref{figure_varying_gamma}.

\subsection{Variance of interparticle distance}

The expression of the covariance can also be used to obtain the variance of the distance between two particles. Indeed
\beq
g_{i,n}^2 = \langle (\delta x_i-\delta x_{i+n})^2 \rangle = \langle \delta x^2_i \rangle+\langle \delta x^2_{i+n} \rangle -2 \langle \delta x_i  \delta x_{i+n} \rangle.
\label{variance_of_gap}
\eeq
One can use the results from the previous subsection to estimate each term in Eq.~\eqref{variance_of_gap}. This leads to
\beq
g_{i,n}^2 =\frac{v^2_0}{N} \mathcal{D}\Big( \frac{x^{\rm{(eq)}}_i}{\sqrt{2N} }, \frac{x^{\rm{(eq)}}_{i+n}}{\sqrt{2N}}\Big),
\label{eq:var_gap_gen_form}
\eeq
where $\mathcal{D}(w,z)=\mathcal{V}(w)+\mathcal{V}(z)- 2 \mathcal{C}(w,z)$ with $\mathcal{C}$ and $\mathcal{V}$ are given in 
Eq.~\eqref{eqn:mathcal_Cb}  and Eq.~\eqref{eqn:mathcal_vb} respectively. 
Let us examine the behavior of this observable near the center of the quadratic well. In the limit where both $w$ and $z$ are close to zero, $\mathcal{D}(w,z)$ admits a simple expression. Indeed in this limit,
\beq
\mathcal{D}(w,z)=\frac{\pi^2}{24}(w-z)^2+\mathcal{O}(w^3,z^3) \;.
\label{eq:db_approximation}
\eeq
Using the fact that near $w=0$, $x^{\rm{(eq)}}_i-x^{\rm{(eq)}}_{i+n} \simeq  \frac{\pi n}{\sqrt{2N}}$. Eqs.~\eqref{eq:var_gap_gen_form} and \eqref{eq:db_approximation} lead to the variance of mid-particle gap ($i=N/2$)
\beq
g^2_{N/2,n}=\langle \Big(\delta x_{N/2} -\delta {x}_{N/2+n} \Big)^2\rangle \simeq \frac{\pi^4}{96} \frac{v^2_0}{ N^3} n^2. 
\label{eq:var_gap_final_form}
\eeq 
\\

\begin{figure}[t]
\includegraphics[width=1.0\linewidth]{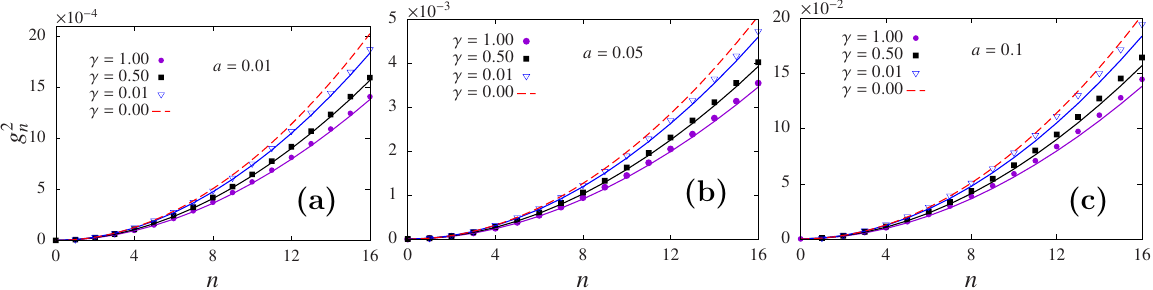}
\caption{Variance of the gap $g^2_n$ as defined in Eq.~\eqref{eq:var_gap_final_form}, is plotted as a function of $n$ for two different values
of the activity parameters, $a=0.01$, $0.05,$ and $0.1$, in panels (a), (b) and (c) respectively, within a system of $128$ particles for three different choices of the tumbling rate $\gamma$. The points represent the numerically computed variance of the gap, while the solid lines indicate the corresponding theoretical expectations at finite $\gamma$ given by Eq.~\eqref{var_chn_gamma} with $i=N/2$. The red dashed line represents the approximated theoretical form of $g^2_n$ given in Eq.~\eqref{eq:var_gap_final_form} obtained in the $\gamma \to 0$ limit. For small values of $n$, the numerical results show good agreement with the theoretical predictions.}
\label{fig:figure_var_gap}
\end{figure}

\noindent{\bf Extension to finite $\gamma$.}
The results in \eqref{covgamma_largeN} and \eqref{vargamma_largeN} enable us to generalize to finite $\gamma$ the expressions in \eqref{eq:var_gap_gen_form} and \eqref{eq:var_gap_final_form}. Indeed,
\begin{eqnarray} \label{var_chn_gamma}
&&g^2_{i,n}=\langle (\delta x_i - \delta x_{i+n})^2 \rangle = \frac{v_0^2}{N} \mathcal{D}^{\gamma} \left(\frac{x^{\rm (eq)}_i}{\sqrt{2N}}, \frac{x^{\rm (eq)}_{i+n}}{\sqrt{2N}}\right) \quad ,~\text{with}~ \\
&&\mathcal{D}^\gamma (w,z) = \mathcal{V}^\gamma(w) + \mathcal{V}^\gamma(z) - 2 \, \mathcal{C}^\gamma(w,z) \nn  \\
&& \ \ \ \ \ \ \ \ \ \ \ = \sum_{k=1}^{\infty} \frac{1}{k^2(k^2+2\gamma)} \left( \frac{\sin(k\arccos w)}{\sqrt{1-w^2}} - \frac{\sin(k\arccos z)}{\sqrt{1-z^2}} \right)^2 \nn
\;.
\end{eqnarray}
The expression \eqref{var_chn_gamma} for the gap variance at finite $\gamma$ is compared with the results of numerical simulations for $N=128$ in Fig.~\ref{fig:figure_var_gap}, and we observe a very good agreement for $a\lesssim 0.1$. We now assume as before that $w,z \ll 1$. Expanding to quadratic order at the center of the trap, the resulting sum over $k$ is still convergent and we find the asymptotic behavior  
\begin{equation} \label{D_asympt_gamma}
\mathcal{D}^\gamma(w,z) \simeq \frac{(w-z)^2}{4} \left( \frac{\pi}{\sqrt{2\gamma}} \coth \big( \pi \sqrt{\frac{\gamma}{2}} \big) -\frac{1}{\gamma} \right) + \mathcal{O}(w^3,z^3) \;,
\end{equation}
which recovers \eqref{eq:db_approximation} for $\gamma\to 0$ (using $\coth(x) = 1/x + x/3 + \mathcal{O}(x^3)$ for $x\to 0$). Here we have used that, if $kw,kz \ll 1$,
\begin{eqnarray}
\!\!\!\!\!\!\!\!\!\!\!\!\!\!\!\! \left(\frac{\sin(k\arccos w)}{\sqrt{1-w^2}} - \frac{\sin(k\arccos z)}{\sqrt{1-z^2}} \right)^2 &\simeq& \begin{cases} (\cos(kw) - \cos(kz))^2 \; , \; k \text{ odd} \\
(\sin(kw) - \sin(kz))^2 \; , \; k \text{ even} \end{cases} \\
&\simeq& \begin{cases} \mathcal{O}(w^4,z^4) \; , \; k \text{ odd} \\
k^2 (w-z)^2 \; , \; k \text{ even} \end{cases} \;.
\end{eqnarray}
Using again that near the center of the trap, $x^{\rm{(eq)}}_i-x^{\rm{(eq)}}_{i+n} \simeq  \frac{\pi n}{\sqrt{2N}}$, we obtain the generalization of \eqref{eq:var_gap_final_form},
\begin{equation}
    g_{N/2,n}^2=\langle (\delta x_{N/2} - \delta x_{N/2+n})^2 \rangle \simeq \frac{\pi^2}{16} \frac{v_0^2}{N^3} \left( \frac{\pi}{\sqrt{2\gamma}} \coth \big( \pi \sqrt{\frac{\gamma}{2}} \big) -\frac{1}{\gamma} \right) n^2.
    \label{gapvariance_largeN_gamma}
\end{equation}

The analytical predictions \eqref{eq:var_gap_gen_form} and \eqref{var_chn_gamma} are compared with numerical results in Fig.~\ref{fig:figure_var_gap}. The agreement is very good for small values of $a=v_0/N$, for all values of $n=\mathcal{O}(1)$. This is in contrast with what was observed in Ref.~\cite{touzo_arxiv_2023} in the case of the active DBM, where the present approximation failed close to $n=1$. In particular, for the active ISM, the variance of mid-gap near the center of the trap is correctly given to leading order by Eq.~\eqref{eq:var_gap_final_form} ($\gamma \to 0$) and \eqref{gapvariance_largeN_gamma} (any $\gamma$) with $n=1$. This is due to the faster convergence of the series defining $\mathcal{C}^{\gamma}(x,y)$ and $\mathcal{V}^{\gamma}(x)$ compared to the active DBM case.

\subsection{Validity of the approximation}

The results \eqref{eq:var_gap_final_form} and \eqref{gapvariance_largeN_gamma} allow us to test {\textit a posteriori} the validity of the small $v_0$ approximation made at the beginning of this section. This requires that the typical variations of the distance between successive particles $g_{i,1}$ is much smaller than the average distance $\Delta_{i,1}$. Thus we get that our results should be valid when the following ratio is small (assuming that $\gamma$ is small and that $i$ is close to $N/2$ to simplify the evaluation),
\begin{equation}  \label{ratio3} 
  \frac{g_{i,1}}{\Delta_{i,1}} \sim \frac{v_0/N^{3/2}}{1/\sqrt{N}} = \frac{v_0}{N} \;.
\end{equation}
Hence we find that the weak noise approximation should always hold as long as $v_0 \ll N$, i.e., in the entire `solid' and `liquid' regimes. Note that this is different from the case of the active DBM, for which this approximation breaks down for $v_0 \sim \sqrt{N}$.

\section{Conclusions and outlook}
\label{section_conclusion}
We extended the analysis of the active Dyson Brownian motion presented in Refs.~\cite{Touzo_EPL_2023,touzo_arxiv_2023} to the active inverse-square power-law model described by Eq.~\eqref{eqn_of_motion}. In the steady state, we first focused on the density profile of the particles. As system activity increases, we observe an interesting change in the density profile: it transitions from a sharply peaked form to a smooth, bell-shaped distribution, with the Wigner semi-circle profile appearing in between. The first crossover connects a crystalline regime, marked by sharp peaks in the density profile, to a liquid-like phase with a smooth Wigner semi-circular profile devoid of such peaks. We have characterized quantitatively this crossover by computing the Lindemann's ratio, which compares the fluctuations in particle positions to the average gaps between successive particles. As observed in the active DBM~\cite{Touzo_EPL_2023}, we also found that the density profile in the steady state retains the shape of a Wigner semi-circle even at large activity. However, at extremely high activity ($v_0\sim\mathcal{O}(N)$), the density profile becomes bell-shaped. To characterize this second crossover, we measured the deviation of the density profile from the Wigner semi-circle by computing the distance measure, $\chi$. Based on these values, we built a `phase' diagram in the $(a,1/\gamma)$ plane where $v_0=a N$ (see Fig.~\ref{fig:phase_diagram_together}).

Using a Hessian approximation, we computed the covariances of particle positions in the small activity (small $v_0$) limit, which also allowed us to obtain the variance of inter-particle gaps. Closed-form expressions for the variance of particle positions and inter-particle gaps were obtained both in the long persistence time limit ($\gamma \to 0$), where they can be expressed in terms of simple functions, and for finite $\gamma$, 
for large system sizes. These theoretical predictions remain valid at sufficiently large $v_0$ (in fact up to $v_0\sim \mathcal{O}(N)$) in both regimes of $\gamma$. 
By comparing the fluctuations of the particle positions with the other relevant length scales in the model, we also identified the crossover regions from weakly active to intermediate active, and finally to strongly active regimes. The first transition from weakly to intermediate active regime occurs at $v_0 \sim \mathcal{O}(1)$, while the transition to the strongly active regime takes place at $v_0 \sim \mathcal{O}(N)$. 

All these observations should be compared with what was found for the active DBM, which was studied in Refs.\cite{Touzo_EPL_2023, touzo_arxiv_2023}. Qualitatively, the behavior of the two models is very similar. In particular, both models have the same `phase' diagram, with a first crossover from a crystal to a liquid regime and a second crossover to a bell-shaped density. Moreover, these crossovers occur for similar values of $v_0$ (note that the Refs.~\cite{Touzo_EPL_2023, touzo_arxiv_2023} use a different scaling of the parameters). The differences between the two models become visible when looking more precisely at the finite $N$ effects in the density and at the fluctuations in the different regimes, in particular at the edges of the support.

This study raises several interesting open questions, which call for further extensions. First, 
it is now evident that, for both the active DBM and the active ISM, the density profile maintains its passive structure, i.e., the Wigner semi-circle, even at sufficiently large activity. It is thus natural to wonder whether a similar behavior occurs in other many-particle interacting systems, such as the harmonically confined Riesz gas~\cite{Agarwal_PRL_2019}. A recent work addressed the properties of the Riesz gas on the circle \cite{Riesz2024}, but does not allow to study the effects related to the presence of an edge in the density. Similarly, one can ask
what happens if one shortens the range of the interactions~\cite{Kumar_PRE_2020, Santra_JPA_2024}.
Another challenging open question would be the description of the bell-shaped density profile, as well as the ubiquitous $1/x^3$ tail, emerging in the large activity limit in both the active DBM and active ISMs.

Finally we comment on the possibilities of experimental  observation of some of our results including the crystal to liquid cross-over. A number of experiments have looked at steady states of one or more trapped active particles in the dilute non-interacting limit \cite{Takatori2016, Dauchot2019, Buttinoni2022}. So far the effect of interactions has not been experimentally probed. We note that active particles are now realized using diverse systems such as bacteria, vibrated granular matter, Janus particles and robots. With current technological capabilities \cite{Murali2022, Paramanick2024}, it may be possible to achieve active particles with  effective power-law interactions. Hence we believe that experimental observation of the melting is possible and would be interesting to explore.

\section{Acknowledgements}
GS and PLD acknowledge support from ANR Grant No. ANR- 23-CE30-0020-01 EDIPS. SS, CD, AD, SD and AK acknowledge support from the Department of Atomic Energy, Government of India, under Project No. RTI4001. AD acknowledges the J.C. Bose Fellowship (JCB/2022/000014) of the Science and Engineering Research Board of the Department of Science and Technology, Government of India. AK would like to acknowledge the support of DST, Government of India Grant under Project No. CRG/2021/002455 and the MATRICS grant MTR/2021/000350 from the SERB, DST, Government of India. PS acknowledges the support of Novo Nordisk Foundation under the Grant Number NNF21OC0071284, for
funding his postdoctoral position.

\appendix

\section{The details of numerical simulation}
\label{numerical_details}
In this Appendix, we provide the details of our numerical simulation. The Euler method~\cite{Atkinson_book} is used to evolve the positions of the particles in the active inverse-square power-law model, following the equations of motion given in Eq.~\eqref{eqn_of_motion}.  To prevent the particles from crossing during the evolution, we chose a time step of $dt \sim 10^{-5}$. The particles are evolved for a long time, $t \sim \mathcal{O}(10^5)$, to ensure that the system has reached a steady state. All physical quantities, such as the density profiles, variance of particle positions, and variance of interparticle gaps in the steady state, are obtained by averaging over a time window of $10^4$ units of time, and over $250$ independent initial conditions which were evolved in parallel. Except for Fig.~\ref{figure_density_large}(a), a bin size of $0.01/\sqrt{N}$ was used to construct the histograms of the density profiles.

\section{The numerical values of the distance measure, $\chi$}
\label{sec:appendixA}

In this Appendix, we provide a table of numerical values of the distance measure $\chi$ for several values of the activity parameter $a$ and the tumbling rate $\gamma$. Recall, $\chi$ measures the difference between the density profile in the steady state with the Wigner semi-circular density profile $\rho_{\rm sc}(x)$~as
\beq
\chi=\int_{-\infty}^{\infty} |\rho(x)-\rho_{\rm sc}(x)| dx,
\eeq
where 
\beq
\rho_{\rm sc}(x)=\begin{cases}
 \frac{1}{\pi N} \sqrt{2N-x^2} ~\text{for}~ |x| \leq \sqrt{2N}\\
 ~~~~~~~0~~~~~~~~~~~~~\text{otherwise}.
 \end{cases}
\eeq
Table \ref{tab:my-table} presents the $\chi$ values for two distinct system sizes, $N=64$ and $128$.
\begin{table}[t]
\centering
\begin{tabular}{|c|c|c|c|}
\hline
Tumbling rate, $\gamma$ & Activity parameter $a$ & $ \chi~\text{for}~N=64$ & $ \chi ~\text{for}~ N=128$ \\ \hline
1.000 & 0.100 & 0.014 & 0.018 \\ \hline
1.000 & 0.250 & 0.084 & 0.084 \\ \hline
1.000 & 0.500 & 0.230 & 0.229 \\ \hline
0.500 & 0.100 & 0.017 & 0.021 \\ \hline
0.500 & 0.250 & 0.099 & 0.096 \\ \hline
0.500 & 0.500 & 0.266 & 0.264 \\ \hline
0.100 & 0.100 & 0.022 & 0.027 \\ \hline
0.100 & 0.250 & 0.122 & 0.117 \\ \hline
0.100 & 0.500 & 0.325 & 0.322 \\ \hline
0.050 & 0.100 & 0.023 & 0.028 \\ \hline
0.050 & 0.250 & 0.126 & 0.120 \\ \hline
0.050 & 0.500 & 0.337 & 0.332 \\ \hline
0.010 & 0.100 & 0.024 & 0.029 \\ \hline
0.010 & 0.250 & 0.130 & 0.125 \\ \hline
0.010 & 0.500 & 0.351 & 0.344 \\ \hline
\end{tabular}
\caption{Table showing the values of the measure $\chi$ which computes the difference between the density profile in the steady state from the Wigner semi-circle profile.}
\label{tab:my-table}
\end{table}
We propose a criterion for the second transition from a smooth Wigner semi-circle to a 
 bell-shaped density profile to occur at $\chi = 0.1$, indicating a $10\%$ deviation in the density profile as compared to $\rho_{\rm sc}(x)$.

\newpage

\section*{References}

\bibliographystyle{unsrt}

\end{document}